\documentclass[10pt,journal,epsfig]{IEEEtran}

\usepackage[dvips]{graphicx}
\usepackage{graphicx}
\usepackage{amssymb}
\usepackage{cite}
\usepackage{subfigure}
\usepackage{amsmath}

\begin{document}

\title{Computationally Efficient Sparse Bayesian Learning via Generalized Approximate Message Passing}

\author{Fuwei Li, Jun Fang, Huiping Duan, Zhi Chen, and Hongbin Li,~\IEEEmembership{Senior
Member,~IEEE}
\thanks{Fuwei Li, Jun Fang, and Zhi Chen are with the National Key Laboratory on Communications,
University of Electronic Science and Technology of China, Chengdu
611731, China, Email: JunFang@uestc.edu.cn}
\thanks{Huiping Duan is with the School of Electronic Engineering,
University of Electronic Science and Technology of China, Chengdu
611731, China, Email: huipingduan@uestc.edu.cn}
\thanks{Hongbin Li is
with the Department of Electrical and Computer Engineering,
Stevens Institute of Technology, Hoboken, NJ 07030, USA, E-mail:
Hongbin.Li@stevens.edu}
\thanks{This work was supported in part by the National Science
Foundation of China under Grant 61172114, and the National Science
Foundation under Grant ECCS-1408182.}}

\maketitle

\begin{abstract}
The sparse Beyesian learning (also referred to as Bayesian
compressed sensing) algorithm is one of the most popular
approaches for sparse signal recovery, and has demonstrated
superior performance in a series of experiments. Nevertheless, the
sparse Bayesian learning algorithm has computational complexity
that grows exponentially with the dimension of the signal, which
hinders its application to many practical problems even with
moderately large data sets. To address this issue, in this paper,
we propose a computationally efficient sparse Bayesian learning
method via the generalized approximate message passing (GAMP)
technique. Specifically, the algorithm is developed within an
expectation-maximization (EM) framework, using GAMP to efficiently
compute an approximation of the posterior distribution of hidden
variables. The hyperparameters associated with the hierarchical
Gaussian prior are learned by iteratively maximizing the
Q-function which is calculated based on the posterior
approximation obtained from the GAMP. Numerical results are
provided to illustrate the computational efficacy and the
effectiveness of the proposed algorithm.
\end{abstract}




\section{Introduction}
Compressed sensing is a recently emerged technique for signal
sampling and data acquisition which enables to recover sparse
signals from much fewer linear measurements
\begin{align}
\boldsymbol{y}=\boldsymbol{Ax}+\boldsymbol{w} \label{data-model}
\end{align}
where $\boldsymbol{A}\in\mathbb{R}^{M\times N}$ is the sampling
matrix with $M\ll N$, $\boldsymbol{x}$ denotes an $N$-dimensional
sparse signal, and $\boldsymbol{w}$ denotes the additive noise.
Such a problem has been extensively studied and a variety of
algorithms, e.g. the orthogonal matching pursuit (OMP) algorithm
\cite{TroppGilbert07}, the basis pursuit (BP) method
\cite{CandesTao05}, and the iterative reweighted $\ell_1$ and
$\ell_2$ algorithms \cite{WipfNagaranjan10}, were proposed.
Besides these methods, another important class of compressed
sensing techniques that have received significant attention are
Bayesian methods, among which sparse Bayesian learning (also
referred to as Bayesian compressed sensing) is considered as one
of the most popular compressed sensing methods. Sparse Bayesian
learning (SBL) was originally proposed by Tipping in his
pioneering work \cite{Tipping01} to address the regression and
classification problems. Later on in \cite{Wipf06,JiXue08}, sparse
Bayesian learning was adapted for sparse signal recovery, and
demonstrated superiority over the greedy methods and the basis
pursuit method in a series of experiments. Despite its superior
performance, a major drawback of the sparse Bayesian learning
method is that it requires to compute an inverse of an $N\times N$
matrix at each iteration, and thus has computational complexity
that grows exponentially with the dimension of the signal. This
high computational cost prohibits its application to many
practical problems with even moderately large data sets.




In this paper, we develop a computationally efficient generalized
approximate message passing (GAMP) algorithm for sparse Bayesian
learning. GAMP, introduced by Donoho \emph{et. al.}
\cite{DonohoMaleki09,DonohoMaleki10} and generalized by Rangan
\cite{Rangan11}, is a newly emerged Bayesian iterative technique
developed in a message passing-based framework for efficiently
computing an approximation of the posterior distribution of
$\boldsymbol{x}$, given a pre-specified prior distribution for
$\boldsymbol{x}$ and a distribution for $\boldsymbol{w}$. In many
expectation-maximization (EM)-based Bayesian methods (including
SBL), the major computational task is to compute the posterior
distribution of the hidden variable $\boldsymbol{x}$. GAMP can
therefore be embedded in the EM framework to provide an
approximation of the true posterior distribution of
$\boldsymbol{x}$, thus resulting in a computationally efficient
algorithm. For example, in \cite{SomSchniter12,VilaSchniter13},
GAMP was used to derive efficient sparse signal recovery
algorithms, with a Markov-tree prior or a Gaussian-mixture prior
placed on the sparse signal. In this work, by resorting to GAMP,
we develop an efficient sparse Bayesian learning method for sparse
signal recovery. Simulation results show that the proposed method
performs similarly as the EM-based sparse Bayesian learning
method, meanwhile achieving a significant computational complexity
reduction. We note that an efficient sparse Bayesian learning
algorithm was developed in \cite{TanLi10} via belief propagation.
The work, however, requires a sparse dictionary $\boldsymbol{A}$
to facilitate the algorithm design, which may not be satisfied in
practical applications.

\section{Overview of Sparse Bayesian Learning}
We first provide a brief review of the sparse Bayesian learning
method. In the sparse Bayesian learning framework, a two-layer
hierarchical prior model was proposed to promote the sparsity of
the solution. In the first layer, $\boldsymbol{x}$ is assigned a
Gaussian prior distribution
\begin{align}
p(\boldsymbol{x}|\boldsymbol{\alpha})=\prod_{n=1}^N
p(x_n|\alpha_n)=\prod_{n=1}^N\mathcal{N}(x_n|0,\alpha_n^{-1})
\label{hm-1}
\end{align}
where $\alpha_n$ is a non-negative hyperparameter controlling the
sparsity of the coefficient $x_n$. The second layer specifies
Gamma distributions as hyperpriors over the hyperparameters
$\{\alpha_n\}$, i.e.
\begin{align}
p(\boldsymbol{\alpha})=\prod_{n=1}^N\text{Gamma}(\alpha_n|a,b)=\prod_{n=1}^N\Gamma^{-1}(a)b^{a}\alpha_{n}^{a-1}e^{-b\alpha_{n}}
\nonumber
\end{align}
where $\Gamma(a)=\int_{0}^{\infty}t^{a-1}e^{-t}dt$ is the Gamma
function. Besides, $\boldsymbol{w}$ is assumed Gaussian noise with
zero mean and covariance matrix $(1/\gamma)\boldsymbol{I}$. We
place a Gamma hyperprior over $\gamma$:
$p(\gamma)=\text{Gamma}(\gamma|c,d)=\Gamma(c)^{-1}d^{c}\gamma^{c-1}e^{-d\gamma}$.

An expectation-maximization (EM) algorithm can be developed for
learning the sparse signal $\boldsymbol{x}$ as well as the
hyperparameters $\{\boldsymbol{\alpha},\gamma\}$. In the EM
formulation, the signal $\boldsymbol{x}$ is treated as hidden
variables, and we iteratively maximize a lower bound on the
posterior probability
$p(\boldsymbol{\alpha},\gamma|\boldsymbol{y})$ (this lower bound
is also referred to as the Q-function). Briefly speaking, the
algorithm alternates between an E-step and a M-step. In the
E-step, we need to compute the posterior distribution of
$\boldsymbol{x}$ conditioned on the observed data and the
estimated hyperparameters, i.e.
\begin{align}
p(\boldsymbol{x}|\boldsymbol{y},\boldsymbol{\alpha}^{(t)},\gamma^{(t)})\propto
p(\boldsymbol{x}|\boldsymbol{\alpha}^{(t)})p(\boldsymbol{y}|\boldsymbol{x},\gamma^{(t)})
\end{align}
It can be readily verified that the posterior
$p(\boldsymbol{x}|\boldsymbol{y},\boldsymbol{\alpha}^{(t)},\gamma^{(t)})$
follows a Gaussian distribution with its mean and covariance
matrix given respectively by
\begin{align}
\boldsymbol{\mu}=&\gamma^{(t)}\boldsymbol{\Phi}\boldsymbol{A}^T\boldsymbol{y}
\nonumber\\
\boldsymbol{\Phi}=&
(\gamma^{(t)}\boldsymbol{A}^T\boldsymbol{A}+\boldsymbol{D})^{-1}
\label{eqn-2}
\end{align}
where
$\boldsymbol{D}\triangleq\text{diag}(\alpha_1^{(t)},\ldots,\alpha_N^{(t)})$.
The Q-function, i.e.
$E_{\boldsymbol{x}|\boldsymbol{y},\boldsymbol{\alpha}^{(t)},\gamma^{(t)}}[\log
p(\boldsymbol{\alpha},\gamma|\boldsymbol{y})]$, can then be
computed, where the operator
$E_{\boldsymbol{x}|\boldsymbol{y},\boldsymbol{\alpha}^{(t)},\gamma^{(t)}}[\cdot]$
denotes the expectation with respect to the posterior distribution
$p(\boldsymbol{x}|\boldsymbol{y},\boldsymbol{\alpha}^{(t)},\gamma^{(t)})$.
In the M-step, we maximize the Q-function with respect to the
hyperparameters $\{\boldsymbol{\alpha},\gamma\}$, which leads to
the following update rules
\begin{align}
\alpha_n^{(t+1)}=&\frac{2a-1}{\langle x_n^2\rangle+2b} \nonumber\\
\gamma^{(t+1)}=& \frac{M+2c-2}{\langle
\|\boldsymbol{y}-\boldsymbol{Ax}\|_2^2\rangle+2d} \nonumber
\end{align}
where $\langle \cdot\rangle$ denotes the expectation with respect
to the posterior distribution
$p(\boldsymbol{x}|\boldsymbol{y},\boldsymbol{\alpha}^{(t)},\gamma^{(t)})$.

It can be seen that the EM algorithm, at each iteration, requires
to update the posterior distribution
$p(\boldsymbol{x}|\boldsymbol{y},\boldsymbol{\alpha}^{(t)},\gamma^{(t)})$,
which involves computing an $N\times N$ matrix inverse. Thus the
EM-based sparse Bayesian learning algorithm has a computational
complexity of order $\mathcal{O}(N^3)$ flops, and therefore is not
suitable for many real-world applications with increasingly large
data sets and unprecedented dimensions. We, in the following, will
develop a computationally efficient sparse Bayesian learning
algorithm via GAMP.




\section{Proposed SBL-GAMP Algorithm}
Generalized approximate message passing (GAMP) is a
very-low-complexity Bayesian iterative technique recently
developed \cite{DonohoMaleki10,Rangan11} for providing an
approximation of the posterior distribution
$p(\boldsymbol{x}|\boldsymbol{y},\boldsymbol{\alpha}^{(t)},\gamma^{(t)})$,
conditioned on that the prior distribution for $\boldsymbol{x}$
the distribution for the additive noise $\boldsymbol{w}$ are
factorizable. It therefore can be naturally embedded within the EM
framework to provide an approximate posterior distribution of
$\boldsymbol{x}$ to replace the true posterior distribution. From
the GAMP's point of view, the hyperparameters
$\{\boldsymbol{\alpha},\gamma\}$ are considered as known and
fixed. The hyperparameters can be updated in the M-step based on
the approximate posterior distribution of $\boldsymbol{x}$.


\subsection{GAMP}
GAMP was developed in a message passing-based framework. By using
central-limit-theorem approximations, the message passing between
variable nodes and factor nodes can be greatly simplified, and the
loopy belief-propagation on the underlying factor graph can be
efficiently performed. In general, in the GAMP algorithm
development, the following two important approximations are
adopted.

Let $\boldsymbol{\theta}\triangleq\{\boldsymbol{\alpha},\gamma\}$
denote the hyperparameters. Firstly, GAMP assumes posterior
independence among hidden variables $\{x_n\}$ and approximates the
true posterior distribution
$p(x_n|\boldsymbol{y},\boldsymbol{\theta})$ by
\begin{align}
\hat{p}(x_n|\boldsymbol{y},\hat{r}_n,\tau_n^r,\boldsymbol{\theta})
=&\frac{p(x_n|\boldsymbol{\theta})\mathcal{N}(x_n|\hat{r}_n,\tau_n^r)}{\int_x
p(x_n|\boldsymbol{\theta})\mathcal{N}(x_n|\hat{r}_n,\tau_n^r)}
\label{eqn-1}
\end{align}
where $\hat{r}_n$ and $\tau_n^r$ are quantities iteratively
updated during the iterative process of the GAMP algorithm, here
we have dropped their explicit dependence on the iteration number
$k$ for simplicity. Substituting (\ref{hm-1}) into (\ref{eqn-1}),
it can be easily verified that the approximate posterior
$\hat{p}(x_n|\boldsymbol{y},\hat{r}_n,\tau_n^r,\boldsymbol{\theta})$
follows a Gaussian distribution with its mean and variance given
respectively as
\begin{align}
\mu_n^x&\triangleq\frac{\hat{r}_n}{1+\alpha_n\tau_n^r} \label{x-post-mean} \\
\phi_n^x&\triangleq\frac{\tau_n^r}{1+\alpha_n\tau_n^r}
\label{x-post-var}
\end{align}
The other approximation is made to the noiseless output
$z_m\triangleq\boldsymbol{a}_m^T\boldsymbol{x}$, where
$\boldsymbol{a}_m^T$ denotes the $m$th row of $\boldsymbol{A}$.
GAMP approximates the true marginal posterior
$p(z_m|\boldsymbol{y},\boldsymbol{\theta})$ by
\begin{align}
\hat{p}(z_m|\boldsymbol{y},\hat{p}_m,\tau_m^p,\boldsymbol{\theta})=\frac{p(y_m|z_m,\boldsymbol{\theta})
\mathcal{N}(z_m|\hat{p}_m,\tau_m^p)}{\int_z
p(y_m|z_m,\boldsymbol{\theta})
\mathcal{N}(z_m|\hat{p}_m,\tau_m^p)}
\end{align}
where $\hat{p}_m$ and $\tau_m^p$ are quantities iteratively
updated during the iterative process of the GAMP algorithm, again
here we dropped their explicit dependence on the iteration number
$k$. Under the additive white Gaussian noise assumption, we have
$p(y_m|z_m,\boldsymbol{\theta})=\mathcal{N}(y_m|z_m,1/\gamma)$.
Thus
$\hat{p}(z_m|\boldsymbol{y},\hat{p}_m,\tau_m^p,\boldsymbol{\theta})$
also follows a Gaussian distribution with its mean and variance
given by
\begin{align}
\mu_m^z\triangleq&\frac{\tau_m^p\gamma
y_m+\hat{p}_m}{1+\gamma\tau_m^p} \label{z-post-mean}
\\
\phi_m^z\triangleq&\frac{\tau_m^p}{1+\gamma\tau_m^p}
\label{z-post-var}
\end{align}

With the above approximations, we can now define the following two
important scalar functions: $g_{\text{in}}(\cdot)$ and
$g_{\text{out}}(\cdot)$ that will be used in the GAMP algorithm.
In the minimum mean-squared error (MMSE) mode, the input scalar
function $g_{\text{in}}(\cdot)$ is simply defined as the posterior
mean $\mu_n^x$ \cite{Rangan11}, i.e.
\begin{align}
g_{\text{in}}(\hat{r}_n,\tau_n^r,
\boldsymbol{\theta})=\mu_n^x=\frac{\hat{r}_n}{1+\alpha_n\tau_n^r}
\end{align}
The scaled partial derivative of $\tau_n^r
g_{\text{in}}(\hat{r}_n,\tau_n^r,\boldsymbol{\theta})$ with
respect to $\hat{r}_n$ is the posterior variance $\phi_n^x$, i.e.
\begin{align}
\tau_n^r\frac{\partial}{\partial\hat{r}_n}g_{\text{in}}(\hat{r}_n,\tau_n^r,\boldsymbol{\theta})
=\phi_n^x=\frac{\tau_n^r}{1+\alpha_n\tau_n^r}
\end{align}
The output scalar function $g_{\text{out}}(\cdot)$ is related to
the posterior mean $\mu_m^z$ as follows
\begin{align}
g_{\text{out}}(\hat{p}_m,\tau_m^p,\boldsymbol{\theta})=\frac{1}{\tau_m^p}(\mu_m^z-\hat{p}_m)=
\frac{1}{\tau_m^p}\bigg(\frac{\tau_m^p\gamma
y_m+\hat{p}_m}{1+\gamma\tau_m^p}-\hat{p}_m\bigg)
\end{align}
The partial derivative of
$g_{\text{out}}(\hat{p}_m,\tau_m^p,\boldsymbol{\theta})$ is
related to the posterior variance $\phi_m^z$ in the following way
\begin{align}
\tau_m^p\frac{\partial}{\partial\hat{p}_m}g_{\text{out}}(\hat{p}_m,\tau_m^p,\boldsymbol{\theta})=
\frac{\phi_m^z-\tau_m^p}{\tau_m^p}=\frac{-\gamma\tau_m^p}{1+\gamma\tau_m^p}
\end{align}
Given definitions of $g_{\text{in}}(\cdot)$ and
$g_{\text{out}}(\cdot)$, the GAMP algorithm can now be summarized
as follows (details of the derivation of the GAMP algorithm can be
found in \cite{Rangan11}), in which $a_{mn}$ denotes the $(m,n)$th
entry of $\boldsymbol{A}$, $\mu_n^x(k)$ and $\phi_n^{x}(k)$ denote
the posterior mean and variance of $x_n$ at iteration $k$,
respectively.

\begin{center}
\textbf{GAMP Algorithm}
\end{center}
\vspace{0cm} \noindent
\begin{tabular}{p{8.7cm}}
\hline Initialization: given $\boldsymbol{\theta}^{(t)}$; set
$k=0$, $\hat{s}_m^{(-1)}=0,\forall m\in\{1,\ldots,M\}$;
$\{\mu_n^x(k)\}_{n=1}^N$ and $\{\phi_n^{x}(k)\}_{n=1}^N$ are
initialized as the mean and variance of the prior distribution. \\
Repeat the following steps until
$\sum_{n}|\mu_n^x(k+1)-\mu_n^x(k)|^2\leq\epsilon$, where
$\epsilon$ is a pre-specified error tolerance. \\
Step 1. $\forall m\in\{1,\ldots,M\}$: \\
$\begin{aligned}
\qquad\qquad\quad \hat{z}_m(k)=&\sum_n a_{mn}\mu_n^x(k)\\
\tau_m^p(k)=&\sum_n a_{mn}^2\phi_n^{x}(k) \\
\hat{p}_m(k)=&\hat{z}_m(k)-\tau_m^p(k)\hat{s}_m(k-1)\\
\end{aligned}$ \\
Step 2. $\forall m\in\{1,\ldots,M\}$: \\
$\begin{aligned} \qquad\qquad\quad
\hat{s}_m(k)=&g_{\text{out}}(\hat{p}_m(k),\tau^p_m(k),\boldsymbol{\theta}^{(t)})\\
\tau^s_m(k)=&-\frac{\partial}{\partial\hat{p}_m}g_{\text{out}}(\hat{p}_m(k),\tau^p_m(k),\boldsymbol{\theta}^{(t)})\\
\end{aligned}$ \\
Step 3. $\forall n\in\{1,\ldots,N\}$: \\
$\begin{aligned} \qquad\qquad\quad
\tau_n^r(k)=&\left(\sum_{m}a_{mn}^2\tau^s_m(k)\right)^{-1} \\
\hat{r}_n(k)=&\mu_n^x(k)+\tau_n^r(k)\sum_{m}a_{mn}\hat{s}_m(k) \\
\end{aligned}$ \\
Step 4. $\forall n\in\{1,\ldots,N\}$: \\
$\begin{aligned} \qquad\qquad\quad
\mu_n^x(k+1)=&g_{\text{in}}(\hat{r}_n(k),\tau_n^r(k),\boldsymbol{\theta}^{(t)})
\\
\phi_n^{x}(k+1)=&\tau_n^r(k)\frac{\partial}{\partial\hat{r}_n}g_{\text{in}}(\hat{r}_n(k),\tau_n^r(k),\boldsymbol{\theta}^{(t)})
\\
\end{aligned}$ \\
Output: $\{\hat{r}_n(k_0),\tau_n^r(k_0)\}$,
$\{\hat{p}_m(k_0),\tau_m^p(k_0)\}$, and
$\{\mu_n^x(k_0+1),\phi_n^{x}(k_0+1)\}$, where $k_0$ stands for the
last iteration.\\
\hline
\end{tabular}

\vspace{0.3cm}

We have now derived an efficient algorithm to generate approximate
posterior distributions for the variables $\boldsymbol{x}$ and
$\boldsymbol{z}\triangleq\boldsymbol{Ax}$. We see that the GAMP
algorithm no longer needs to compute an inverse of a matrix. The
dominating operations in each iteration is the simple matrix
multiplications, which scale as $\mathcal{O}(MN)$. Thus the
computational complexity can be significantly reduced. In the
following, we discuss how to update the hyperparameters via the
EM.


\subsection{Hyperparameter Learning via EM}
As indicated earlier, in the EM framework, the hyperparameters are
estimated by treating $\boldsymbol{x}$ as hidden variables and
iteratively maximizing the Q-function, i.e.
\begin{align}
\boldsymbol{\theta}^{(t+1)}=\arg\max_{\boldsymbol{\theta}}
Q(\boldsymbol{\theta}|\boldsymbol{\theta}^{(t)})\triangleq
E_{\boldsymbol{x}|\boldsymbol{y},\boldsymbol{\theta}^{(t)}}[\log
p(\boldsymbol{\theta}|\boldsymbol{x},\boldsymbol{y})]
\end{align}
We first carry out the M-step for the hyperparameters
$\{\alpha_n\}$. We take the partial derivative of the Q-function
with respect to $\alpha_n$, which yields
\begin{align}
\frac{\partial}{\partial\alpha_n}Q(\boldsymbol{\theta}|\boldsymbol{\theta}^{(t)})
=&\frac{\partial}{\partial\alpha_n}E_{\boldsymbol{x}|\boldsymbol{y},\boldsymbol{\theta}^{(t)}}[\log
p(\boldsymbol{\theta}|\boldsymbol{x},\boldsymbol{y})] \nonumber\\
=&\frac{\partial}{\partial\alpha_n}E_{\boldsymbol{x}|\boldsymbol{y},\boldsymbol{\theta}^{(t)}}[\log
p(x_n|\alpha_n)p(\alpha_n; a,b)] \nonumber\\
=&\frac{1}{2\alpha_n}-\frac{\langle
x_n^2\rangle}{2}+\frac{a-1}{\alpha_n}-b \label{eqn-3}
\end{align}
where $\langle\cdot\rangle$ denotes the expectation with respect
to $p(\boldsymbol{x}|\boldsymbol{y},\boldsymbol{\theta}^{(t)})$.
Since the true posterior is unavailable, we use
$\hat{p}(x_n|\boldsymbol{y},\hat{r}_n(k_0),\tau_n^r(k_0),\boldsymbol{\theta}^{(t)})$,
i.e. the approximate posterior distribution of $x_n$ obtained from
the GAMP algorithm to replace the true posterior distribution.
Recalling that
$\hat{p}(x_n|\boldsymbol{y},\hat{r}_n(k_0),\tau_n^r(k_0),\boldsymbol{\theta}^{(t)})$
follows a Gaussian distribution with its mean and variance given
by (\ref{x-post-mean})--(\ref{x-post-var}), we have
\begin{align}
\langle x_n^2\rangle=
\frac{(\hat{r}_n(k_0))^2}{(1+\alpha_n^{(t)}\tau_n^r(k_0))^2}+\frac{\tau_n^r(k_0)}{1+\alpha_n^{(t)}\tau_n^r(k_0)}
\end{align}
Setting (\ref{eqn-3}) equal to zero gives the update rule for
$\alpha_n$
\begin{align}
\alpha_n^{(t+1)}=\frac{2a-1}{2b+\langle x_n^2\rangle} \quad
\forall n\in\{1,\ldots,N\} \label{alpha-update}
\end{align}

We now discuss the update of the hyperparameter $\gamma$, the
inverse of the noise variance. Since the GAMP algorithm also
provides an approximate posterior distribution for the noiseless
output $\boldsymbol{z}$, we can simply treat $\boldsymbol{z}$ as
hidden variables when learning the noise variance, i.e.
\begin{align}
\gamma^{(t+1)}=\arg\max_{\gamma}E_{\boldsymbol{z}|\boldsymbol{y},\boldsymbol{\theta}^{(t)}}[\log
p(\boldsymbol{y}|\boldsymbol{z},\gamma)p(\gamma;c,d)]
\end{align}
Taking the partial derivative of the Q-function with respect to
$\gamma$ gives
\begin{align}
&\frac{\partial}{\partial\gamma}E_{\boldsymbol{z}|\boldsymbol{y},\boldsymbol{\theta}^{(t)}}[\log
p(\boldsymbol{y}|\boldsymbol{z},\gamma)p(\gamma;c,d)] \nonumber\\
=&
\frac{M}{2\gamma}-\frac{1}{2}\sum_{m=1}^M\langle(y_m-z_m)^2\rangle+\frac{c-1}{\gamma}-d
\end{align}
where $\langle\cdot\rangle$ denotes the expectation with respect
to
$p(z_m|\boldsymbol{y},\hat{p}_m(k_0),\tau_m^p(k_0),\boldsymbol{\theta}^{(t)})$,
i.e. the approximate posterior distribution of $z_m$. Recalling
that the approximate posterior of $z_m$ follows a Gaussian
distribution with its mean and variance given by
(\ref{z-post-mean})--(\ref{z-post-var}), we have
\begin{align}
\langle(y_m-z_m)^2\rangle=(y_m-\mu^z_m)^2+\phi_m^z
\end{align}
where $\mu^z_m$ and $\phi_m^z$ are given by
(\ref{z-post-mean})--(\ref{z-post-var}), with
$\{\hat{p}_m,\tau_m^p\}$ replaced by
$\{\hat{p}_m(k_0),\tau_m^p(k_0)\}$, and $\gamma$ replaced by
$\gamma^{(t)}$. Setting the derivative equal to zero, we obtain
the update rule for $\gamma$ as
\begin{align}
\gamma^{(t+1)}=\frac{M+2c-2}{2d+\sum_{m}\langle(y_m-z_m)^2\rangle}
\label{gamma-update}
\end{align}

So far we have completed the development of our GAMP-based sparse
Bayesian learning algorithm. For clarify, we now summarize our
proposed SBL-GAMP algorithm as follows.

\begin{center}
\textbf{SBL-GAMP Algorithm}
\end{center}
\vspace{0cm} \noindent
\begin{tabular}{lp{7.7cm}}
\hline 1. &Initialization: given $\boldsymbol{\alpha}^{(0)}$ and $\gamma^{(0)}$.\\
2. &For $t\geq 0$: given $\boldsymbol{\alpha}^{(t)}$ and
$\gamma^{(t)}$, recall the GAMP algorithm. Based on the outputs of
the GAMP algorithm, update the hyperparameters
$\boldsymbol{\alpha}^{(t+1)}$ and
$\gamma^{(t+1)}$ according to (\ref{alpha-update}) and (\ref{gamma-update}).\\
3. & Continue the above iteration until the difference between two
consecutive estimates of $\boldsymbol{x}$
is negligible.\\
\hline
\end{tabular}


\section{Simulation Results}
We now carry out experiments to illustrate the performance of the
proposed SBL-GAMP algorithm\footnote{Codes are available at
http://www.junfang-uestc.net/codes/SBL-GAMP.rar}. In our
simulations, the $K$-sparse signal is randomly generated with its
support set randomly chosen according to a uniform distribution.
The measurement matrix $\boldsymbol{A}\in\mathbb{R}^{M\times N}$
is randomly generated with each entry independently drawn from
Gaussian distribution with zero mean and unit variance, and then
each column of $\boldsymbol{A}$ is normalized to unit norm. We
compare our method with the conventional EM-based sparse Bayesian
learning (referred to as SBL-EM) method \cite{Tipping01} and the
BP-AMP algorithm \cite{DonohoMaleki10}.

\begin{figure}[!t]
 \centering
\begin{tabular}{cc}
\hspace*{-3ex}
\includegraphics[width=4.9cm,height=4.9cm]{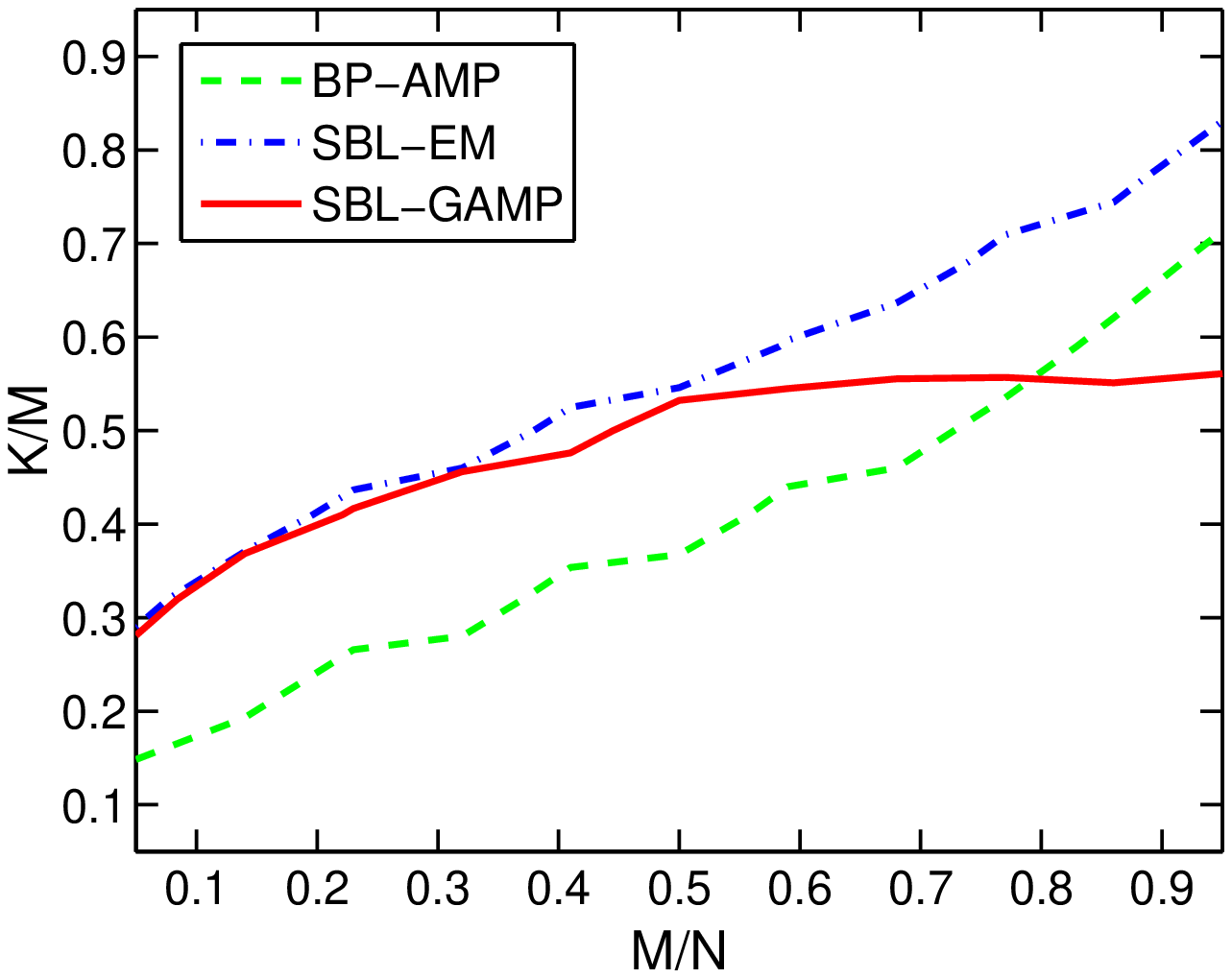}&
\hspace*{-5ex}
\includegraphics[width=4.9cm,height=4.9cm]{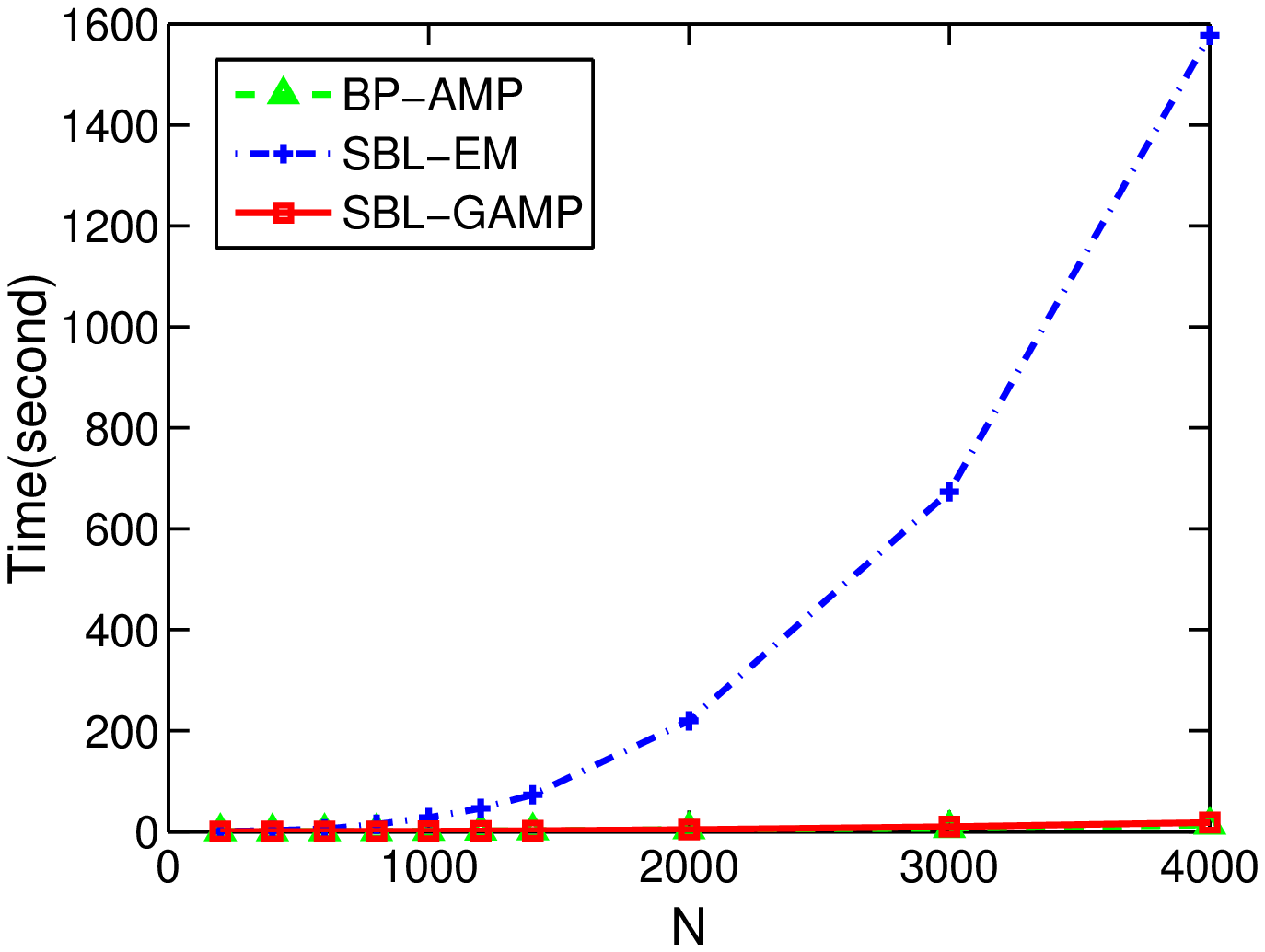}
\\
(a)& (b)
\end{tabular}
  \caption{(a). Phase transitions of respective algorithms; (b). Average run times vs. $N$.}
   \label{fig1}
\end{figure}

\begin{figure}[!t]
\centering
\includegraphics[width=7cm]{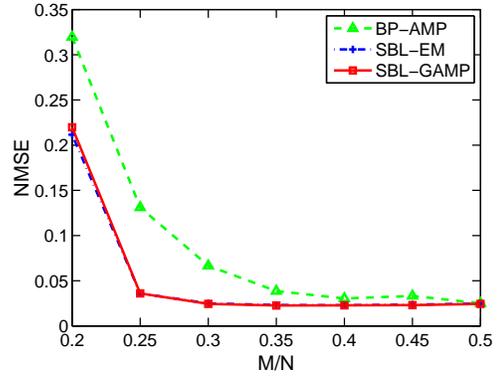}
\caption{NMSEs of respective algorithms vs. the ratio $M/N$.}
\label{fig2}
\end{figure}


We first examine the phase transition behavior of respective
algorithms. The phase transition is used to illustrate how
sparsity level ($K/M$) and the oversampling ratio ($M/N$) affect
the success rate of each algorithm in exactly recovering sparse
signals in noiseless scenarios. In particular, each point on the
phase transition curve corresponds to a success rate equal to
$0.5$. The success rate is computed as the ratio of the number of
successful trials to the total number of independent runs. A trial
is considered successful if the normalized squared error
$\|\boldsymbol{x}-\boldsymbol{\hat{x}}\|_2^2/\|\boldsymbol{x}\|_2^2$
is no greater than $10^{-6}$. Fig. \ref{fig1}(a) plots the phase
transitions of respective algorithms, where we set $N=1000$, and
the oversampling ratio $M/N$ varies from $0.05$ to $0.95$. From
Fig. \ref{fig1}(a), we see that, when $M/N<0.5$, the proposed
SBL-GAMP algorithm achieves performance similar to SBL-EM, and is
superior to BP-AMP. The proposed method is surpassed by BP-AMP as
the oversampling ratio increases. Nevertheless, SBL-GAMP is still
more appealing since we usually prefer compressed sensing
algorithms work under high compression rate regions. The average
run times of respective algorithms as a function of the signal
dimension $N$ is plotted in Fig. \ref{fig1}(b), where we set
$M=0.4N$ and $K=0.3M$. Results are averaged over 10 independent
runs. We see that the SBL-GAMP consumes much less time than the
SBL-EM due to its easy computation of the posterior distribution
of $\boldsymbol{x}$, particularly for a large signal dimension
$N$. Also, it can be observed that the average run time of the
SBL-EM grows exponentially with $N$, whereas the average run time
of the SBL-GAMP grows very slowly with an increasing $N$. This
observation coincides with our computational complexity analysis
very well. Lastly, we examine the recovery performance in a noisy
scenario, where we set $N=500$, $K=40$, and the signal to noise
ratio (SNR) is set to 20dB. Fig. \ref{fig2} depicts the normalized
mean square errors (NMSE) of respective algorithms vs. $M/N$.
Results are averaged over 1000 independent runs. We see that the
SBL-GAMP algorithm achieves a similar recovery accuracy as the
SBL-EM algorithm even with a much lower computational complexity.


\section{Conclusions}
We developed a computationally efficient sparse Bayesian learning
(SBL) algorithm via the GAMP technique. The proposed method has a
much lower computational complexity (of order $\mathcal{O}(MN)$)
than the conventional SBL method. Simulation results show that the
proposed method achieves recovery performance similar to the
conventional SBL method in the low oversampling ratio regime.


\end{document}